%% file: main.tex
\documentclass[10pt,twocolumn,letterpaper]{article}

\usepackage[pagenumbers]{cvpr} 

\input{preamble}

\definecolor{cvprblue}{rgb}{0.21,0.49,0.74}
\usepackage[pagebackref,breaklinks,colorlinks,allcolors=cvprblue]{hyperref}

\def\confName{CVPR}
\def\confYear{2027}

\title{Perception Over Geometry: Rethinking Gaussian Allocation in Mesh-based Splatting}

\author{
Yuting Wang\textsuperscript{1}\quad \quad
Yang Zhao\textsuperscript{1,3,$\dagger$}\quad \quad
Shanding Diao\textsuperscript{1}\quad \quad
Penglin Wang\textsuperscript{1}
\\[3pt]
Yuan Chen\textsuperscript{4}\quad \quad
Wei Jia\textsuperscript{1}\quad \quad
Ronggang Wang\textsuperscript{2,3}
\\[6pt]
\textsuperscript{1}Hefei University of Technology \quad
\textsuperscript{2}Peking University Shenzhen Graduate School
\\[3pt]
\textsuperscript{3}Peng Cheng Laboratory \quad
\textsuperscript{4}Anhui University
}

\begin{document}
\input{sec/0_abstract}    
\input{sec/1_intro}

\input{sec/2_relatedwork}

\input{sec/3_method}

\input{sec/4_experiments}

{
    \small
    \bibliographystyle{ieeenat_fullname}
    \bibliography{main}
}
\input{sec/X_suppl}

\end{document}

%% file: preamble.tex
\usepackage{cuted} 

\usepackage{currfile} 

\usepackage{caption} 

\usepackage{tabularx}  
\usepackage{booktabs}  
\usepackage{multirow}  
\usepackage{array}     



%% file: sec/0_abstract.tex
\twocolumn[{%
    \renewcommand\twocolumn[1][]{#1}%
    \maketitle

    \centering
    \includegraphics[width=\textwidth]{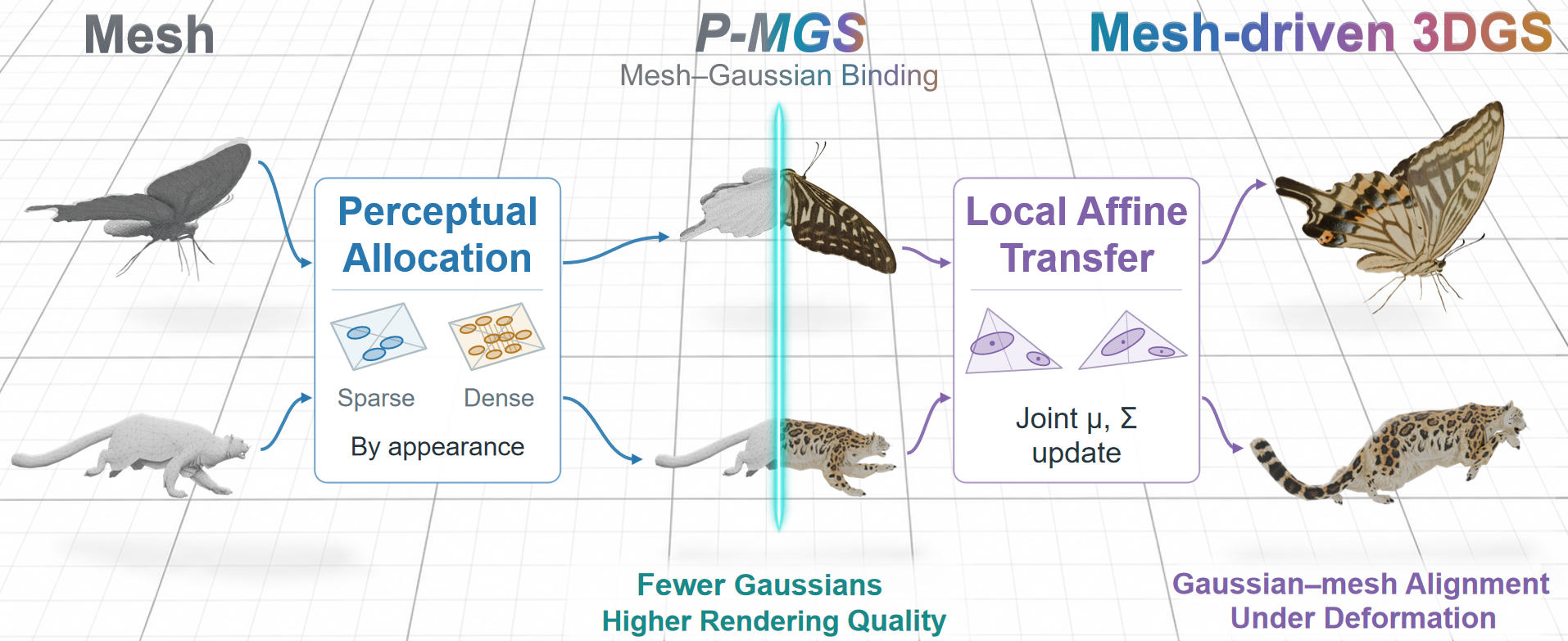}

    {
    \captionsetup{hypcap=false}
    \captionof{figure}{
            P-MGS enables high-quality Gaussian rendering with flexible mesh-driven editing.
    More visual results and source code are available on our
    \href{https://github.com/yuting01234/P-MGS}{project page}.
    }
    \label{fig:abstruct}
    }

    \vspace{1em}
}]

\begin{abstract}
3D Gaussian Splatting (3DGS) enables photorealistic novel-view synthesis but lacks explicit topology for geometric editing. Existing mesh-based methods enable geometric manipulation by associating Gaussians with triangular mesh faces, typically assigning a fixed number of Gaussians to each face. However, the representation capacity required for appearance reconstruction depends on local appearance complexity rather than merely on mesh tessellation density. 
To address these limitations, we propose P-MGS, a framework that separates appearance modeling from geometric control to achieve high quality rendering, compact representation, and flexible mesh editing. P-MGS introduces a perceptual complexity guided allocation strategy that distributes Gaussians according to surface appearance demand estimated from multiview observations. To preserve reconstruction flexibility, Gaussians are optimized without mesh constraints and subsequently associated with the original mesh through post hoc binding, retaining adaptive refinement while preserving mesh topology. During mesh editing, local affine deformation transfer updates Gaussian centers and covariances under mesh deformation, reducing misalignment and floating artifacts. Experiments demonstrate superior average novel view synthesis quality with fewer Gaussians and coherent mesh editing. 
\end{abstract}

%% file: sec/1_intro.tex
\begin{figure*}[t]
    \centering
    \includegraphics[width=0.9\textwidth]{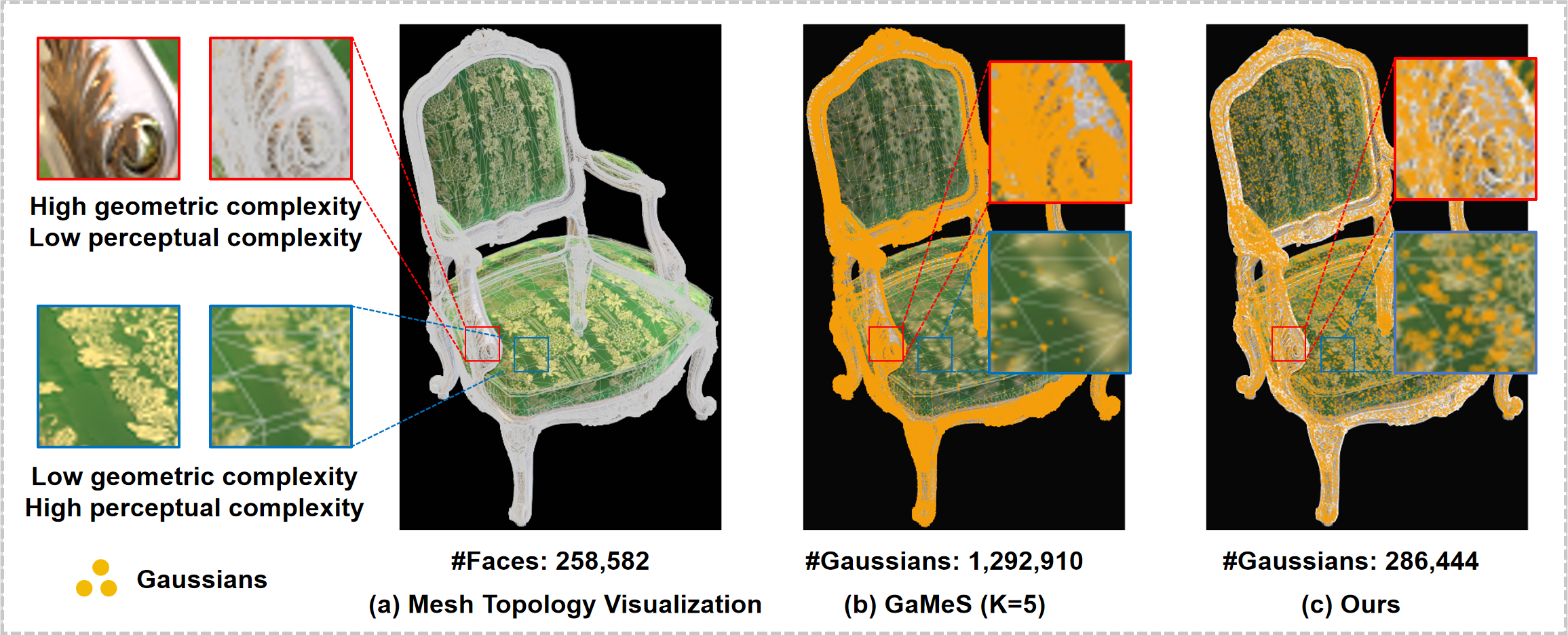}
    \caption{
    Illustration of the mismatch between mesh subdivision density and perceptual complexity.
    (a) Mesh topology with non-uniform geometric complexity.
    (b) Fixed Gaussian allocation in GaMeS~\cite{games}.
    (c) Perceptual complexity-guided Gaussian allocation in P-MGS.
    }
    \label{fig:observation}
\end{figure*}

\section{Introduction}
\label{sec:intro}

3D scene reconstruction is fundamental to computer vision and graphics, supporting applications such as embodied intelligence, robotic manipulation, and digital twins. Recently, 3D Gaussian Splatting (3DGS) has emerged as a powerful representation for real-time photorealistic rendering and novel-view synthesis~\cite{3dgs}. Through differentiable splatting of anisotropic Gaussian primitives, 3DGS achieves efficient optimization and high-quality rendering. However, Gaussian primitives are independently optimized without explicit geometric structures or surface connectivity, making structure-aware manipulation and deformation consistency challenging~\cite{sugar,games,manigs}.

In contrast, triangle meshes provide explicit surface topology and geometric connectivity for editing, deformation, and simulation~\cite{nerfediting,neumesh,neuralimpostor}. However, mesh-based representations are limited by surface discretization, making it difficult to simultaneously preserve geometric structures and detailed appearance. 
Therefore, combining Gaussian Splatting with triangle meshes provides a promising direction, where Gaussians capture appearance while meshes provide geometric controllability~\cite{sugar,games,gaussianfrosting,manigs}.

Among mesh-based Gaussian Splatting methods, GaMeS~\cite{games} enables mesh-driven deformation by binding Gaussians to mesh triangles. It assigns $K$ Gaussians to each of the $N_f$ faces, yielding $N_G=N_f \times K$ primitives.
However, this fixed allocation assumes that mesh tessellation reflects appearance complexity. As illustrated in Fig.~\ref{fig:observation}, densely tessellated yet visually simple regions receive redundant Gaussians, whereas texture-rich regions with coarse tessellation remain underrepresented.
Moreover, this mesh--Gaussian correspondence is established during optimization, where Gaussian parameters are constrained by mesh structures~\cite{sugar,games,gaomesh,manigs}. Such a tightly coupled optimization process limits Gaussian flexibility and prevents leveraging the adaptive optimization capability of vanilla 3DGS~\cite{3dgs}. During mesh-driven editing, inconsistent deformation of Gaussian positions and shapes can introduce Gaussian--mesh misalignment and floating artifacts, particularly under non-rigid transformations.

This work presents P-MGS, a mesh--Gaussian splatting framework that combines perceptual complexity-guided allocation with decoupled reconstruction and manipulation. Guided by multi-view appearance, P-MGS allocates Gaussians according to perceptual complexity rather than mesh tessellation, improving representation efficiency. Gaussians are initialized with mesh guidance, optimized without mesh constraints, and subsequently associated with the mesh through post-hoc binding. 
During editing, local affine transfer updates Gaussian centers and covariances according to mesh transformations, maintaining alignment and reducing floating artifacts under non-rigid deformation.

Our main contributions are summarized as follows:
\begin{itemize}

\item We rethink the coupling between mesh geometry and Gaussian representation through a \emph{Perception Over Geometry} principle, where meshes provide structural controllability while appearance complexity governs Gaussian allocation. Based on this principle, we develop a multi-view perceptual allocation strategy that adapts representation capacity to visual demand.

\item We develop a decoupled reconstruction and manipulation framework, enabling unconstrained Gaussian optimization followed by post-hoc Gaussian--mesh binding. For editing, we introduce local affine deformation transfer to maintain Gaussian--mesh alignment under deformation.

\item Extensive experiments on NeRF-Synthetic~\cite{mildenhall} and DTU~\cite{dtu} demonstrate high-quality novel-view synthesis with compact Gaussian representations and flexible mesh-driven editing.

\end{itemize}

%% file: sec/2_relatedwork.tex
\section{Related Work}
\label{sec:relatedwork}

\subsection{Traditional Mesh-based Representations}
Traditional mesh-based representations have been used in image-based rendering due to their explicit geometric structures and compatibility with graphics pipelines~\cite{mobilenerf}. 
Conventional methods reconstruct textured meshes from multi-view images using structure-from-motion and multi-view stereo~\cite{colmap}, followed by texture mapping and rasterization for novel-view synthesis. Despite efficient rendering, their fidelity depends on geometric reconstruction accuracy and texture resolution, limiting the preservation of fine geometry and high-frequency appearance details.

Neural and point-based scene representations offer alternatives to conventional mesh rendering~\cite{neus,nerfshop}. 
Neural Radiance Fields (NeRF)~\cite{mildenhall} model scenes as continuous volumetric functions, enabling high-quality novel-view synthesis. 
Subsequent methods improve rendering efficiency through sparse representations and explicit data structures~\cite{nerfacc,plenoxels}. In parallel, 3D Gaussian Splatting (3DGS)~\cite{3dgs} represents scenes with anisotropic Gaussian primitives optimized through differentiable splatting, enabling real-time photorealistic rendering. 
Extensions address anti-aliasing~\cite{yu}, dynamic scene reconstruction~\cite{wu,scgs}, and relighting~\cite{jiang}. However, the lack of explicit topology limits geometric editing, deformation, and physics-based interaction~\cite{scgs,physgaussian,vrgs}.

\subsection{Mesh-Gaussian Hybrid Representations}

Existing mesh-based Gaussian methods commonly initialize Gaussian primitives by sampling points on mesh faces, often assigning a fixed number of Gaussians per triangle and establishing explicit Gaussian--mesh correspondence. Methods such as SuGaR~\cite{sugar}, GaMeS~\cite{games}, GaussianMesh~\cite{gaomesh}, and Mani-GS~\cite{manigs} adopt variants of this mesh-guided paradigm. While providing geometric correspondence~\cite{gaussianfrosting}, these representations tie Gaussian capacity to mesh subdivision rather than appearance-driven rendering demand. Mesh tessellation determines geometric discretization, whereas multi-view observations reveal the spatial distribution of visual complexity. Regions with stronger image-space variations demand greater representation capacity regardless of mesh density. Consequently, uniform face-based allocation overallocates Gaussians to geometrically dense but visually simple regions while undersampling areas with rich textures or high-frequency details. This mismatch motivates allocating Gaussians according to perceptual complexity rather than mesh subdivision alone.

Beyond Gaussian allocation, mesh-driven editing requires consistent Gaussian deformation. GaMeS~\cite{games} parameterizes Gaussian centers on triangle surfaces and derives covariances from triangle-aligned orientations and scales, while Mani-GS~\cite{manigs} supports off-surface Gaussians through local coordinates and edge-length-based scaling. Neither employs a unified local affine transformation to propagate Gaussian centers and covariances, leaving their geometric consistency under anisotropic stretching and shearing unguaranteed. This motivates a unified affine transfer that consistently transforms Gaussian centers and covariances with the mesh.
Furthermore, establishing Gaussian--mesh correspondence before optimization couples Gaussian learning with mesh geometry and restricts the mesh-independent adaptive refinement of vanilla 3DGS. Although GaMeS~\cite{games} introduces a mesh-free Triangle Soup representation, its mesh-based formulation and Mani-GS~\cite{manigs} rely on mesh-conditioned Gaussian optimization. GaussianMesh~\cite{gaomesh} performs adaptive Gaussian refinement through mesh face subdivision, thereby modifying the original mesh tessellation. These approaches motivate decoupling unconstrained Gaussian reconstruction from binding to an editable mesh. 
UniMGS~\cite{unimgs} also associates optimized Gaussians with proxy meshes for deformation, but primarily focuses on unified mesh--Gaussian rasterization and manipulation. In contrast, our work emphasizes appearance-driven Gaussian allocation combined with unconstrained optimization and post-hoc binding.

%% file: sec/3_method.tex
\begin{figure*}[t]
    \centering
    \includegraphics[width=0.95\linewidth]{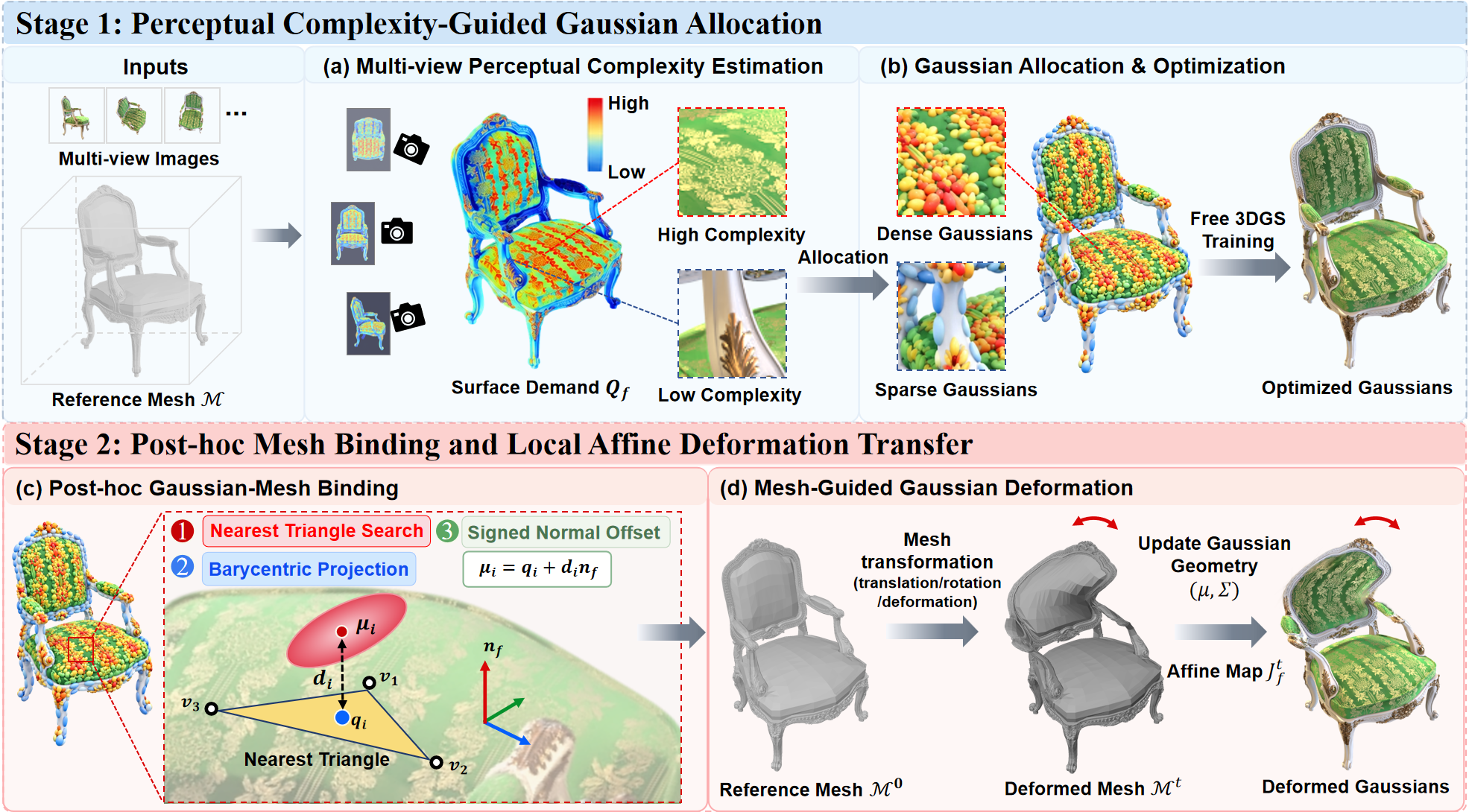}
    \caption{
    Overview of P-MGS.
    The framework consists of two stages: perceptual complexity-guided Gaussian allocation and post-hoc Gaussian--mesh binding.
    In the first stage, multi-view perceptual complexity guides mesh-based Gaussian initialization, followed by unconstrained 3DGS optimization.
    In the second stage, post-hoc binding establishes Gaussian--mesh correspondence, enabling mesh-driven deformation through local affine transformations.
    }
    \label{fig:framework}
    \vspace{-15pt}
\end{figure*}

\section{Method}
\label{sec:method}
Given calibrated multi-view images and a reference mesh, P-MGS decouples Gaussian reconstruction from mesh-driven editing, as illustrated in Fig.~\ref{fig:framework}.
Visible coverage and appearance variation guide Gaussian initialization, followed by unconstrained 3DGS optimization (Sec.~\ref{sec:perceptual_allocation}).
After reconstruction, post-hoc binding establishes Gaussian--mesh correspondence, while local affine deformation transfer maintains alignment between Gaussian primitives and the deformed mesh during editing (Sec.~\ref{sec:Binding}).

\subsection{Preliminary}
\textbf{3D Gaussian Splatting.}
3DGS~\cite{3dgs} represents scenes as explicit anisotropic Gaussian primitives, enabling efficient novel-view synthesis through differentiable splatting. Each primitive $\mathcal{G}_i=\{\boldsymbol{\mu}_i,\boldsymbol{r}_i,\boldsymbol{s}_i,\alpha_i,\mathbf{c}_i\}$ has learnable center, rotation, scale, opacity, and appearance attributes, respectively. Its spatial distribution is given by
\begin{equation}
G_i(\mathbf{x})
=
\exp\left(
-\frac{1}{2}
(\mathbf{x}-\boldsymbol{\mu}_i)^T
\boldsymbol{\Sigma}_i^{-1}
(\mathbf{x}-\boldsymbol{\mu}_i)
\right),
\end{equation}
where $\boldsymbol{\Sigma}_i$ controls the spatial extent and orientation and is parameterized as $\boldsymbol{\Sigma}_i=R_iS_iS_i^TR_i^T$, using rotation matrix $R_i$ and scaling matrix $S_i$.

During rendering, each Gaussian is projected onto the image plane, producing a 2D footprint $G_i^{2D}(\mathbf{p})$ with effective pixel alpha $\widetilde{\alpha}_i(\mathbf{p})=\alpha_iG_i^{2D}(\mathbf{p})$. The projected Gaussians are sorted front to back by camera-space depth and alpha-composited as
\begin{equation}
\mathbf{C}(\mathbf{p})
=
\sum_{i\in\mathcal{N}(\mathbf{p})}
T_i(\mathbf{p})\widetilde{\alpha}_i(\mathbf{p})\mathbf{c}_i,
\label{eq:alpha_compositing}
\end{equation}
where $\mathcal{N}(\mathbf{p})$ denotes the ordered set of contributing Gaussians and $T_i(\mathbf{p})=\prod_{j<i}(1-\widetilde{\alpha}_j(\mathbf{p}))$ represents their accumulated transmittance. Here, $\mathbf{c}_i$ is the RGB color evaluated from the appearance parameters for the current view. The Gaussian attributes are optimized by minimizing the difference between rendered images and captured views.

\textbf{Mesh-guided Gaussian Representation.}
GaMeS~\cite{games} associates Gaussian primitives with triangular mesh faces. Given a face $f_j=(\mathbf{v}_1,\mathbf{v}_2,\mathbf{v}_3)$, the center $\boldsymbol{\mu}_i$ of an associated Gaussian is parameterized by learnable barycentric coordinates:
\begin{equation}
\boldsymbol{\mu}_i
=
\alpha_{i,1}\mathbf{v}_1
+
\alpha_{i,2}\mathbf{v}_2
+
\alpha_{i,3}\mathbf{v}_3 ,
\end{equation}
where $\alpha_{i,1}$, $\alpha_{i,2}$, and $\alpha_{i,3}$ satisfy $\alpha_{i,1}+\alpha_{i,2}+\alpha_{i,3}=1$. This parameterization establishes explicit Gaussian--mesh correspondence, allowing Gaussian primitives to follow mesh surface changes during geometric manipulation.


\subsection{Perceptual Complexity-Guided Gaussian Allocation}
\label{sec:perceptual_allocation}

Existing mesh-based Gaussian representations exploit mesh structures as geometric priors to establish Gaussian--surface correspondence~\cite{games,gaomesh,manigs}.
They commonly initialize the same number of primitives on each triangular face, $n_f=K$, where $n_f$ denotes the initial Gaussian count on face $f$ and $K$ is a constant. 
Although this strategy naturally preserves geometric correspondence, it implicitly assumes that geometric subdivision reflects the representation capacity required for appearance reconstruction.
However, geometry and appearance describe different properties of a surface: mesh topology determines spatial discretization, while multi-view observations determine the perceptual information density required for rendering.

We therefore replace a uniform count per face with an observation-guided allocation under a shared initialization budget.
For each face, we derive a demand score $Q_f$ from an approximation-error model and use it to balance reconstruction capacity against initialization workload.

\subsubsection{Multi-view Perceptual Complexity Estimation}

Let $\mathcal{M}$ be the reference mesh with face set $\mathcal{F}$, and let $\{I_v\}_{v=1}^{V}$ be the training images.
For pixel $p$ in view $v$, we define $h_{fv}(p)=\chi_{fv}(p)a_v(p)$, where $\chi_{fv}(p)$ indicates whether the camera ray first intersects face $f$ and $a_v(p)\in[0,1]$ is the foreground alpha.
Local appearance variation is measured by $\rho_v(p)=\frac{1}{3}\sum_{c\in\{R,G,B\}}\|\nabla I_{v,c}(p)\|_2^2+\epsilon$, where $\nabla$ denotes the spatial gradient operator in the image plane, RGB intensities are normalized to $[0,1]$, and $\epsilon>0$.

For each face-view pair, the visible projected coverage is
$A_{fv}=\sum_p h_{fv}(p)$.
The mean gradient energy is
$\bar{\rho}_{fv}=\sum_p h_{fv}(p)\rho_v(p)/A_{fv}$,
set to zero if $A_{fv}=0$.
To relate these observations to the initial count $n_f$,
we approximate the visible region by $n_f$ equal-area cells
of comparable shape.
For a locally linear image signal, the integrated squared error
of a constant-valued cell scales with gradient energy and squared cell area.
Summing over the cells gives an error proportional to
$A_{fv}^{2}\bar{\rho}_{fv}/n_f$.
Averaging across views yields the demand coefficient
\begin{equation}
Q_f=\frac{1}{V}\sum_{v=1}^{V}A_{fv}^{2}\bar{\rho}_{fv}.
\label{eq:multiview_demand}
\end{equation}
Fig.~\ref{fig:framework}(a) shows that faces with visible coverage or stronger image variation have larger demand scores.
The squared coverage term accounts for cell size in the approximation model, while gradient energy captures appearance variation.
For positive-demand faces, the model gives a multi-view error proportional to $Q_f/n_f$.
This cell-based error serves as a surrogate for allocating initialization capacity.

\subsubsection{Perceptual-aware Gaussian Allocation}

Given a shared initialization budget $B$, we determine the face counts by minimizing the demand-based error surrogate while limiting the workload in each training view.
Let $\mathcal{P}=\{f\in\mathcal{F}:Q_f>0\}$ denote the positive-demand faces, and let $\ell_f$ and $u_f$ be the face-level count bounds.
Let $T_{fv}$ denote the estimated number of image tiles touched by a Gaussian on face $f$ in view $v$, and let $W_v>0$ be the per-view workload limit.
Because projected footprints differ, the same total Gaussian count can incur different rasterization workloads.
The allocation problem is
\begin{equation}
\begin{aligned}
\min_{\mathbf{n}\in\mathbb{Z}_{\geq0}^{|\mathcal{F}|}}
&\quad \mathcal{E}(\mathbf{n})=\sum_{f\in\mathcal{P}}\frac{Q_f}{n_f}\\
\text{s.t.}
&\quad \sum_f n_f=B,\quad \ell_f\leq n_f\leq u_f,\ \forall f,\\
&\quad \sum_f n_fT_{fv}\leq W_v,\ \forall v.
\end{aligned}
\label{eq:constrained_allocation}
\end{equation}
The lower bounds retain at least one initial primitive on each eligible face, and the upper bounds limit local counts using visible-pixel observations.
Positive-demand faces have $\ell_f\geq1$, while eligible zero-demand faces retain only the minimum.
The workload limits are calibrated from a uniform reference initialization.

We approximately solve Eq.~\ref{eq:constrained_allocation} through Lagrangian workload pricing.
For nonnegative view prices $\lambda_v$, increasing a face's count from $k$ to $k+1$ reduces its surrogate error by $Q_f/[k(k+1)]$ and adds a normalized workload $T_{fv}/W_v$ in each view.
The resulting priced marginal benefit is
\begin{equation}
S_f(k)=\frac{Q_f}{k(k+1)}
-\sum_{v=1}^{V}\lambda_v\frac{T_{fv}}{W_v},
\qquad f\in\mathcal{P},\ k\geq1.
\label{eq:priced_allocation_gain}
\end{equation}
Each increment is thus ranked by its error reduction after subtracting the priced workload cost.
The allocation favors faces where an additional Gaussian provides a larger error reduction at a lower priced cost.
Larger prices penalize faces that are expensive in the corresponding views.

For each price evaluation, we reserve the lower-bound counts and distribute the remaining budget by selecting the largest available marginal benefits, subject to the upper bounds, until the total count reaches $B$.
We then update the view prices using normalized workload residuals and recompute the allocation.
Among the workload-feasible candidates encountered, we retain the one with the lowest $\mathcal{E}$.

As shown in Fig.~\ref{fig:framework}(b), we use the allocated counts to sample Gaussian centers on their assigned faces and initialize covariance orientations and scales from local geometry.
After initialization, all Gaussian parameters are optimized freely with the standard 3DGS pipeline, including adaptive densification.
The budget and workload constraints apply to initialization only.
Bound definitions, workload calibration, and solver details are provided in Appendix~\ref{sec:allocation_implementation}.

\subsection{Post-hoc Mesh Binding and Local Affine Deformation Transfer}
\label{sec:Binding}
Unconstrained reconstruction allows Gaussian primitives to move independently of their initialization faces, which improves representation flexibility but removes explicit mesh correspondence. 
We therefore establish Gaussian--mesh binding after optimization, preserving reconstruction freedom while enabling mesh-driven deformation.

\begin{table*}[t]
\centering
\small
\caption{
Quantitative comparison on the NeRF-Synthetic dataset.
For baselines with mesh-based initialization, we assign five Gaussians per face.
GaMeS uses an input mesh, while GaMeS$_{\triangle}$ denotes
the mesh-free Triangle Soup variant.
The best results are marked in \textbf{bold}, and the second-best results are underlined.
}
\label{tab:psnr}

\renewcommand{\arraystretch}{0.95}
\setlength{\tabcolsep}{3pt}

\begin{tabularx}{\textwidth}{@{}lc*{9}{>{\centering\arraybackslash}X}@{}}
\toprule
\multirow{2}{*}{\textbf{Method}} &
\multirow{2}{*}{\textbf{Metric}} &
\multicolumn{8}{c}{\textbf{Scene}} &
\multirow{2}{*}{\textbf{Avg}} \\
\cmidrule(lr){3-10}
& &
Chair & Drums & Ficus & Hotdog &
Lego & Materials & Mic & Ship & \\
\midrule

\multirow{2}{*}{\shortstack[l]{3DGS\\TOG 2023}}
& PSNR $\uparrow$
& \textbf{35.82} & \underline{26.17} & 34.86
& \underline{37.69} & \underline{35.78}
& \underline{30.25} & 36.38 & 30.88 & 33.48 \\
& \#GS $\downarrow$
& \underline{300K} & \underline{550K} & \underline{350K}
& 350K & \underline{1200K} & \underline{800K}
& \underline{500K} & \underline{1000K} & \underline{631K} \\
\midrule

\multirow{2}{*}{\shortstack[l]{SuGaR\\CVPR 2024}}
& PSNR $\uparrow$
& 34.23 & 25.78 & 32.09 & 36.53 & 35.05
& 28.65 & 35.39 & 28.55 & 32.03 \\
& \#GS $\downarrow$
& 1293K & 5675K & 922K & 330K & 10176K
& 6855K & 3994K & 5773K & 4377K \\
\midrule

\multirow{2}{*}{\shortstack[l]{Mani-GS\\CVPR 2025}}
& PSNR $\uparrow$
& 35.37 & 26.13 & \underline{35.24}
& 37.41 & 35.49 & 28.15
& \underline{37.18} & 30.99 & 33.25 \\
& \#GS $\downarrow$
& 917K & 3405K & 553K & 546K & 6106K
& 4112K & 2396K & 3464K & 2687K \\
\midrule

\multirow{2}{*}{\shortstack[l]{MeshSplatting\\CVPR 2026}}
& PSNR $\uparrow$
& 30.75 & 25.04 & 30.34 & 33.09 & 30.95
& 26.33 & 30.12 & 28.29 & 29.36 \\
& \#GS $\downarrow$
& - & - & - & - & - & - & - & - & - \\
\midrule

\multirow{2}{*}{\shortstack[l]{GaMeS\\PR 2027}}
& PSNR $\uparrow$
& 30.70 & 25.35 & 32.28 & 33.40 & 33.20
& 25.30 & 32.47 & 29.31 & 30.25 \\
& \#GS $\downarrow$
& 1293K & 5675K & 922K & \underline{330K} & 10176K
& 6855K & 3994K & 5773K & 4377K \\
\midrule

\multirow{2}{*}{\shortstack[l]{GaMeS$_{\triangle}$\\PR 2027}}
& PSNR $\uparrow$
& \underline{35.57} & \underline{26.17} & 34.81
& 37.65 & 35.42 & 30.01
& 36.82 & \underline{31.50}
& \underline{33.49} \\
& \#GS $\downarrow$
& 608K & 1873K & 523K & 436K & 2764K
& 1230K & 920K & \textbf{920K} & 1159K \\
\midrule

\multirow{2}{*}{Ours}
& PSNR $\uparrow$
& \textbf{35.82} & \textbf{26.31}
& \textbf{35.31} & \textbf{38.07}
& \textbf{36.76} & \textbf{30.47}
& \textbf{37.40} & \textbf{31.82}
& \textbf{34.00} \\
& \#GS $\downarrow$
& \textbf{286K} & \textbf{503K}
& \textbf{333K} & \textbf{156K}
& \textbf{1106K} & \textbf{643K}
& \textbf{414K} & \textbf{920K}
& \textbf{545K} \\
\bottomrule
\end{tabularx}
\vspace{-10pt}
\end{table*}

\textbf{Reference binding.}
Given the reference mesh $\mathcal{M}^{0}$, let $f_i\in\mathcal{F}$ denote the reference face assigned to the $i$-th optimized Gaussian $\mathcal{G}_i$.
We determine this assignment using the spatial matching rule described in Appendix~\ref{sec:binding_implementation}.
Rather than anchoring the Gaussian to a single face vertex, we express its center $\boldsymbol{\mu}_i^0$ using barycentric coordinates and a signed normal offset:
\begin{equation}
\boldsymbol{\mu}_i^0
=
\sum_{k=1}^{3} w_{ik}\mathbf{v}_{f_i,k}^{0}
+
d_i\mathbf{n}_{f_i}^{0},
\label{eq:reference_gaussian_binding}
\end{equation}
where $\mathbf{v}_{f_i,k}^{0}$ and $\mathbf{n}_{f_i}^{0}$ denote the face vertices and unit normal, respectively.
The barycentric weights $w_{ik}$, satisfying $\sum_k w_{ik}=1$, are computed by projecting the Gaussian center onto the face plane, while $d_i$ preserves its signed distance from that plane.
This decomposition retains the optimized Gaussian position without forcing it onto the mesh surface.
The face assignment, barycentric weights, and normal offset are fixed after binding.

\textbf{Local affine deformation transfer.}
For an edited mesh $\mathcal{M}^{t}$ with unchanged topology, the stored barycentric coordinates and normal offsets allow Gaussian centers to follow their associated triangles. To preserve Gaussian--mesh alignment under non-rigid deformation, we construct a face-wise affine map that transfers local stretching, shearing, and rotation to Gaussian shapes.
For $a\in\{0,t\}$, define
\begin{equation}
\mathbf{E}_f^a =
\left[
\mathbf{v}_{f,2}^a-\mathbf{v}_{f,1}^a,\;
\mathbf{v}_{f,3}^a-\mathbf{v}_{f,1}^a,\;
\mathbf{n}_f^a
\right],
\quad
\mathbf{J}_f^t =
\mathbf{E}_f^t(\mathbf{E}_f^0)^{-1}.
\label{eq:local_affine_map}
\end{equation}
The two edge vectors capture in-plane deformation, while the unit normal completes the local 3D frame without introducing additional normal scaling.
We then update the Gaussian center and covariance as
\begin{align}
\boldsymbol{\mu}_i^t
&=
\sum_{k=1}^{3}w_{ik}\mathbf{v}_{f_i,k}^{t}
+
d_i\mathbf{n}_{f_i}^{t},
\label{eq:deformed_gaussian_center}\\
\boldsymbol{\Sigma}_i^t
&=
\mathbf{J}_{f_i}^t
\boldsymbol{\Sigma}_i^0
(\mathbf{J}_{f_i}^t)^\top.
\label{eq:deformed_gaussian_covariance}
\end{align}
Both updates follow the same local affine transformation: Gaussian centers track the deformed mesh faces, while their covariances adapt to local stretching, shearing, and rotation. This coordinated deformation maintains Gaussian--mesh alignment and reduces floating artifacts during non-rigid editing. Under rigid mesh motion, the updates reduce to a rigid transformation.

After binding, mesh edits determine Gaussian deformation without further optimization.
In our experiments, we use Blender to manipulate the reference meshes and demonstrate mesh-driven rendering under large-scale, local, articulated, and non-rigid deformations.
Additional binding and implementation details are provided in Appendix~\ref{sec:binding_affine_details}.

%% file: sec/4_experiments.tex
\section{Experiments}
\label{sec:experiments}



\subsection{Datasets, Metrics, and Baselines}
\label{sec:Data}
We evaluate P-MGS on the NeRF-Synthetic dataset~\cite{mildenhall} and the DTU dataset~\cite{dtu} containing real-world objects. 
For methods requiring mesh inputs, we extract explicit reference meshes using NeuS~\cite{neus}.
We assess rendering quality using PSNR, SSIM~\cite{ssim}, and LPIPS~\cite{lpips}, and report the number of Gaussian primitives to measure representation size.
We compare P-MGS with vanilla 3DGS~\cite{3dgs} and mesh-based Gaussian methods, including SuGaR~\cite{sugar}, GaMeS~\cite{games}, and Mani-GS~\cite{manigs}. We include MeshSplatting~\cite{meshsplatting} as an editable rendering baseline that supports direct geometric manipulation. 
Since it does not use Gaussian primitives, its Gaussian count is marked as ``--''. 
We evaluate GaMeS with mesh-based initialization using input meshes, while GaMeS$_{\triangle}$ denotes the mesh-free Triangle Soup variant.

\begin{figure*}[t]
    \centering
    \includegraphics[width=\textwidth]{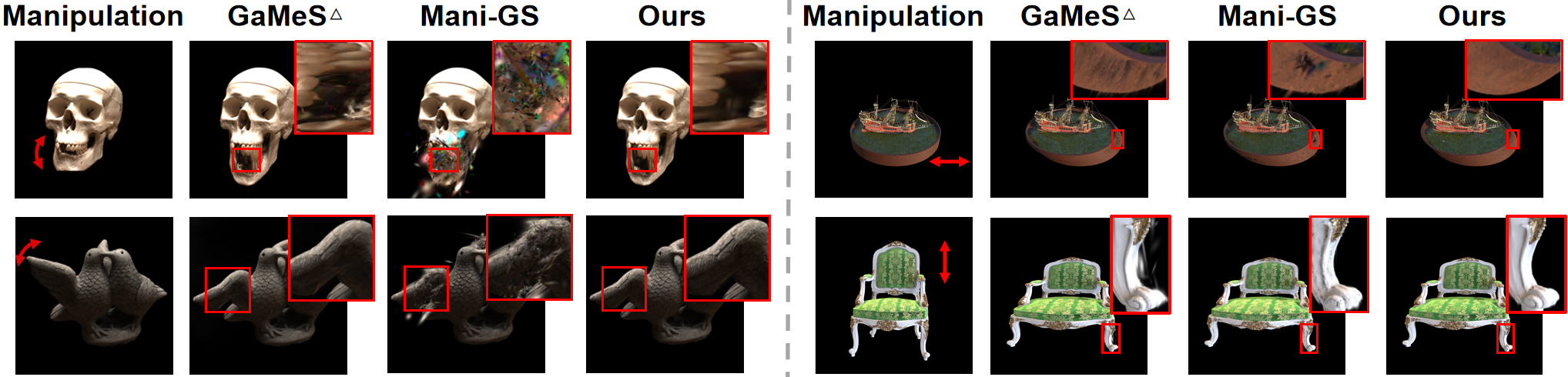}
\caption{
    Qualitative comparison of mesh-driven editing with GaMeS$_{\triangle}$ and Mani-GS on \textit{Chair} and \textit{Ship} from NeRF-Synthetic, and \textit{Pigeon} (Scan106) and \textit{Skull} (Scan65) from DTU.
    GaMeS$_{\triangle}$ and Mani-GS exhibit floating Gaussian artifacts and Gaussian--mesh misalignment after deformation, whereas P-MGS produces fewer artifacts.
}
    \label{fig:manipulation_rendering}
    \vspace{-10pt}
\end{figure*}

\subsection{Evaluation}
\textbf{Static Rendering.}
Table~\ref{tab:psnr} presents the quantitative comparison of novel-view synthesis on the NeRF-Synthetic dataset.
P-MGS achieves the highest average PSNR of 34.00 dB while using the fewest Gaussians (545K) among all Gaussian-based methods.
Compared with Mani-GS, P-MGS improves PSNR by 0.75 dB while reducing the Gaussian count by 79.7\%.
P-MGS also outperforms vanilla 3DGS in both rendering quality and representation efficiency, while additionally supporting mesh-driven editing.
These results demonstrate that P-MGS achieves superior rendering quality with a compact Gaussian representation while preserving mesh-driven editability.

\begin{figure}[t]
    \centering
    \includegraphics[width=\columnwidth]{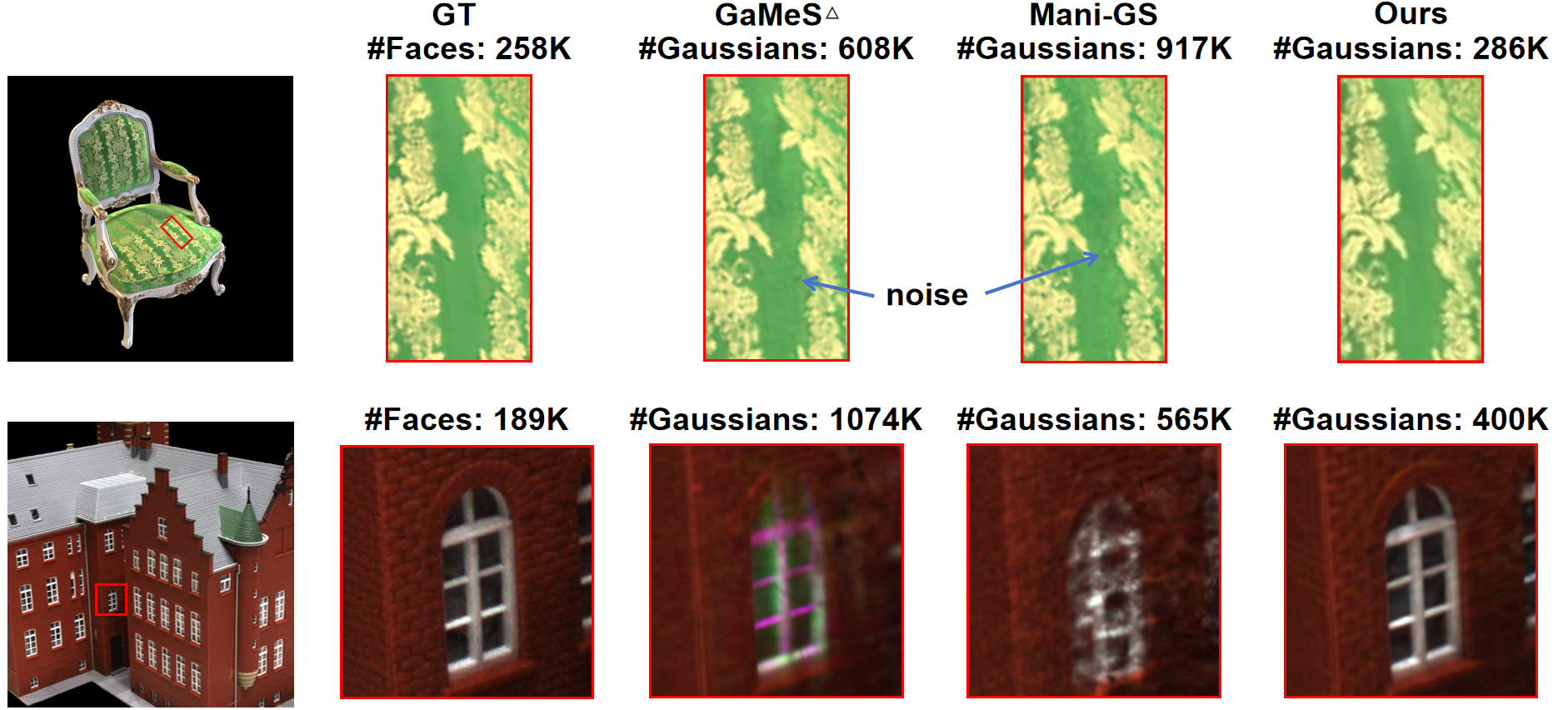}
    \caption{
    Qualitative comparison of static novel-view rendering on \textit{Chair} from NeRF-Synthetic and \textit{Building} (Scan24) from DTU.
    The zoomed-in regions highlight the green area of \textit{Chair} and the windows of \textit{Building}, where P-MGS exhibits fewer noise artifacts than GaMeS$_{\triangle}$ and Mani-GS.
    }
    \label{fig:static_rendering}
    \vspace{-15pt}
\end{figure}

Interestingly, the scene-wise PSNR gaps between unconstrained 3DGS and GaMeS (Mesh-Init) are consistent with a mismatch between geometric and appearance complexity. On \textit{Chair} and \textit{Hotdog}, where geometric complexity does not fully reflect the richness of surface appearance, GaMeS (Mesh-Init) trails 3DGS by 5.12 dB and 4.29 dB, respectively. Increasing representation size alone does not necessarily overcome this limitation: on \textit{Lego}, GaMeS (Mesh-Init) uses 10176K Gaussians---approximately $8.5\times$ the count of 3DGS---yet achieves a PSNR that is 2.58 dB lower. By contrast, \textit{Drums} and \textit{Ship} contain substantial geometric detail but comparatively simpler appearance patterns, and the corresponding gaps are smaller, at 0.82 dB and 1.57 dB, respectively. These observations highlight a limitation of fixed per-face allocation: representation capacity is governed by mesh tessellation, which need not reflect local appearance demand. This motivates allocating Gaussians according to observed appearance complexity while preserving the flexibility of unconstrained optimization.

\begin{table}[t]
\centering
\small
\caption{
Quantitative rendering quality comparison with GaMeS$_{\triangle}$ and Mani-GS on the DTU dataset.}
\label{tab:dtu}

\renewcommand{\arraystretch}{1.05}
\setlength{\tabcolsep}{3pt}

\begin{tabularx}{\columnwidth}{@{}l*{4}{>{\centering\arraybackslash}X}@{}}
\toprule
\textbf{Method}
& \textbf{PSNR $\uparrow$}
& \textbf{SSIM $\uparrow$}
& \textbf{LPIPS $\downarrow$}
& \textbf{\#GS $\downarrow$}
\\
\midrule

GaMeS$_{\triangle}$
& \underline{30.91}
& 0.8732
& \underline{0.0995}
& 603K
\\

Mani-GS
& 30.79
& \underline{0.9067}
& 0.1295
& \underline{316K}
\\

Ours
& \textbf{32.36}
& \textbf{0.9416}
& \textbf{0.0943}
& \textbf{191K}
\\

\bottomrule
\end{tabularx}
\vspace{-15pt}
\end{table}

We evaluate P-MGS on real-world objects from the DTU dataset, as shown in Table~\ref{tab:dtu}. P-MGS outperforms GaMeS$_{\triangle}$ and Mani-GS across all rendering quality metrics while using fewer Gaussians. These results demonstrate the effectiveness of our method in achieving high-fidelity rendering with a compact Gaussian representation.

Figure~\ref{fig:static_rendering} compares static renderings on \textit{Chair} and \textit{Building}.
On \textit{Chair}, P-MGS produces smoother, more coherent green fabric while preserving texture details and reducing the granular noise observed in GaMeS$_{\triangle}$ and Mani-GS.
On \textit{Building}, GaMeS$_{\triangle}$ exhibits false colors within windows, while Mani-GS produces blurred, locally discontinuous window frames with noise.
P-MGS preserves window colors and structural boundaries, yielding closer agreement with the ground truth.
These results demonstrate improved artifact suppression and appearance detail preservation.

\textbf{Manipulation Rendering.}
We evaluate mesh-driven editing using static multi-view datasets with manually applied mesh deformations. Since ground-truth images of the deformed states are unavailable, PSNR cannot be directly evaluated after manipulation.
Figure~\ref{fig:manipulation_rendering} presents qualitative comparisons across different object categories and deformation types. 
Across all examples, GaMeS$_{\triangle}$ and Mani-GS exhibit noticeable floating Gaussian artifacts or Gaussian--mesh misalignment after deformation. These issues are particularly evident in the edited regions, where the rendered appearance becomes locally inconsistent and structural boundaries are degraded. In contrast, P-MGS maintains better Gaussian--mesh alignment, resulting in more coherent object shapes, sharper boundaries, and cleaner appearance details while preserving rendering quality under both large-scale and local manipulations.

\subsection{Ablation Study}

\begin{figure}[t]
    \centering
    \includegraphics[width=0.95\columnwidth]{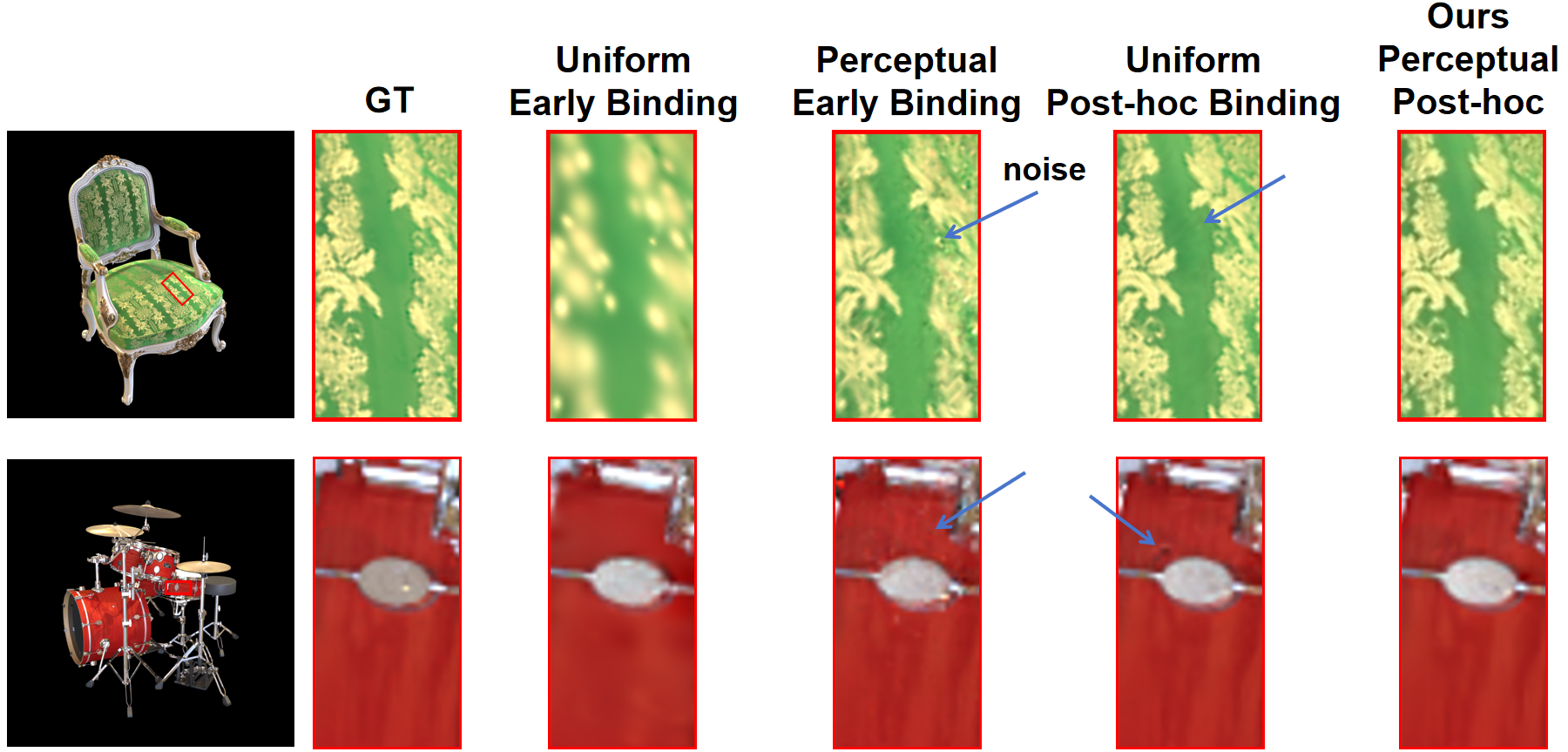}
    \caption{
    Qualitative ablation of Gaussian allocation and
    Gaussian--mesh binding on the \textit{Chair} and \textit{Drums} scenes.
    }
    \label{fig:xiaorong}
    \vspace{-15pt}
\end{figure}

We evaluate perceptual Gaussian allocation and the decoupled optimization strategy with post-hoc binding
against uniform allocation and early binding with mesh-constrained optimization. Table~\ref{tab:ablation} shows that perceptual allocation improves PSNR from 30.10 to 31.36 dB under early binding and from 33.29 to 34.00 dB under post-hoc binding. Unlike early binding, the decoupled strategy permits unconstrained appearance optimization before establishing Gaussian--mesh correspondence for editing. Combining perceptual allocation and post-hoc binding
achieves the best results: 34.00 dB PSNR, 0.971 SSIM, and 0.0279 LPIPS. Figure~\ref{fig:xiaorong} illustrates sharper appearance details and cleaner renderings, highlighting the complementary benefits of both design choices. These results support decoupling appearance learning from geometric control for editing.

\begin{table}[t]
\centering
\footnotesize
\caption{
Ablation of Gaussian allocation and Gaussian--mesh binding
strategies on the NeRF-Synthetic dataset.
Early binding employs mesh-constrained optimization,
whereas post-hoc binding follows unconstrained optimization.
}
\label{tab:ablation}

\renewcommand{\arraystretch}{1.12}
\setlength{\tabcolsep}{2pt}

\begin{tabularx}{\columnwidth}{
@{}ll*{3}{>{\centering\arraybackslash}X}@{}}
\toprule
\textbf{Allocation}
& \shortstack[l]{\textbf{Binding}\\\textbf{Strategy}}
& \textbf{PSNR $\uparrow$}
& \textbf{SSIM $\uparrow$}
& \textbf{LPIPS $\downarrow$}
\\
\midrule

Uniform ($K=3$)
& Early binding
& 30.10
& 0.950
& 0.0476
\\

Uniform ($K=3$)
& Post-hoc binding
& 33.29
& 0.962
& 0.0346
\\

Perceptual
& Early binding
& 31.36
& 0.956
& 0.0436
\\

\textbf{Perceptual}
& \textbf{Post-hoc binding}
& \textbf{34.00}
& \textbf{0.971}
& \textbf{0.0279}
\\

\bottomrule
\end{tabularx}
\vspace{-10pt}
\end{table}

To evaluate perceptual complexity-guided Gaussian allocation, we compare it with uniform allocation using three and five Gaussians per face under mesh-constrained optimization. Table~\ref{xiaorong2} shows that perceptual allocation achieves 31.36 dB PSNR with 2626K Gaussians, outperforming uniform allocation ($K=3$) by 1.26 dB under an identical Gaussian budget. Compared with uniform allocation ($K=5$), our strategy improves PSNR by 1.11 dB while using 40\% fewer Gaussians. Improvements in SSIM and LPIPS demonstrate the effectiveness of perceptual complexity in guiding Gaussian allocation. These results highlight how perceptual allocation concentrates representation capacity where it contributes most to rendering quality.

\begin{table}[t]
\centering
\small
\caption{
Comparison of Gaussian allocation strategies under
mesh-constrained optimization on NeRF-Synthetic.
Perceptual allocation achieves higher rendering quality
with the same Gaussian budget as uniform allocation
($K=3$), while using 40\% fewer Gaussians than
uniform allocation ($K=5$).
}
\label{xiaorong2}

\renewcommand{\arraystretch}{1.05}
\setlength{\tabcolsep}{3pt}

\begin{tabularx}{\columnwidth}{
@{}l*{4}{>{\centering\arraybackslash}X}@{}}
\toprule
\textbf{Allocation}
& \textbf{\#GS $\downarrow$}
& \textbf{PSNR $\uparrow$}
& \textbf{SSIM $\uparrow$}
& \textbf{LPIPS $\downarrow$}
\\
\midrule

Uniform ($K=3$)
& \textbf{2626K}
& 30.10
& 0.950
& 0.0476
\\

Uniform ($K=5$)
& 4377K
& 30.25
& 0.951
& 0.0454
\\

\textbf{Perceptual}
& \textbf{2626K}
& \textbf{31.36}
& \textbf{0.956}
& \textbf{0.0436}
\\

\bottomrule
\end{tabularx}
\vspace{-10pt}
\end{table}

\begin{figure}[t]
    \centering
    \includegraphics[width=0.95\columnwidth]{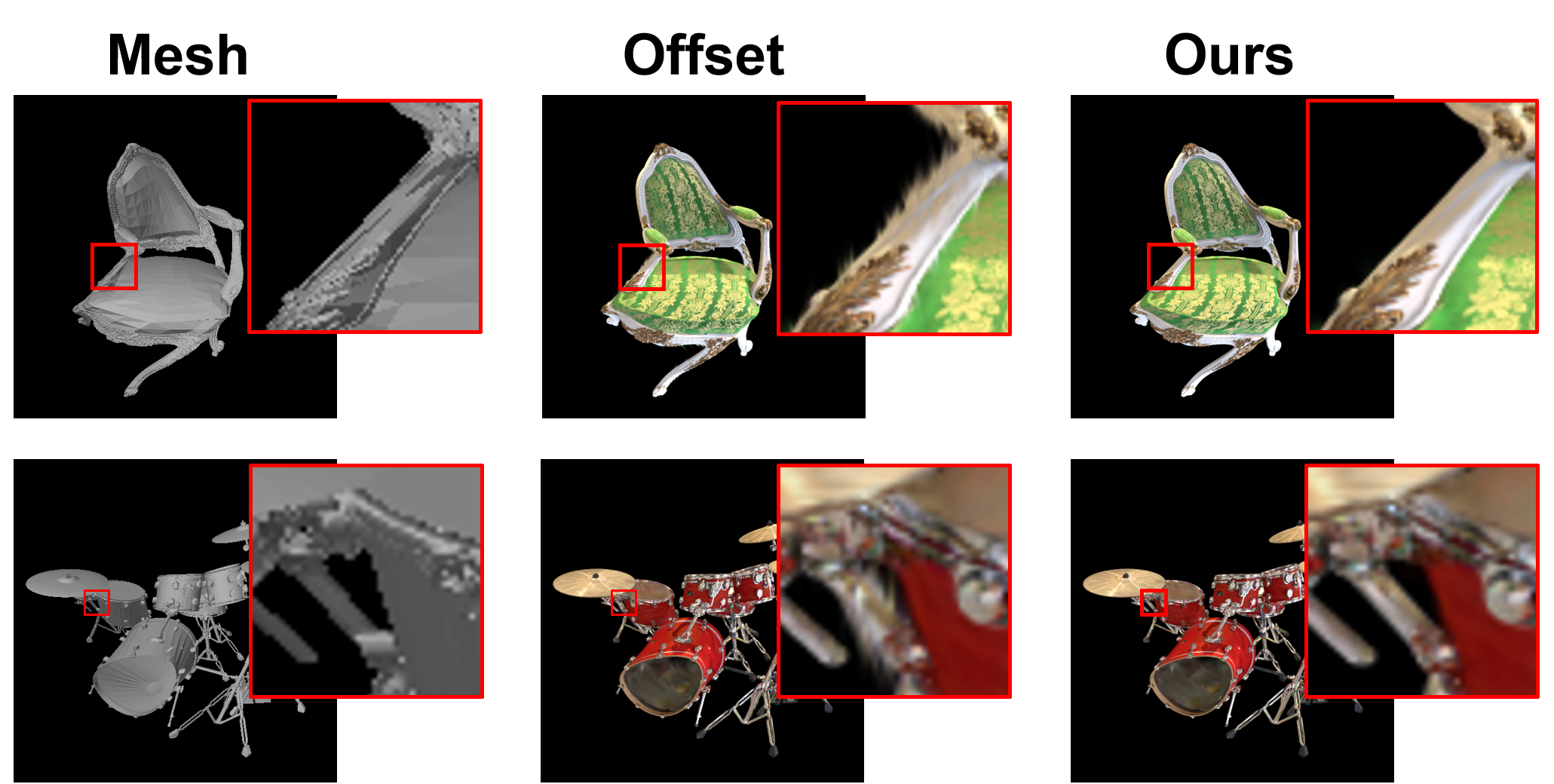}
    \caption{
Qualitative ablation study of mesh-driven deformation on \textit{Chair} and \textit{Drums}. Compared with offset binding and unbound Gaussians, P-MGS maintains Gaussian--mesh alignment and produces cleaner rendering results under non-rigid deformations.
    }
    \label{fig:dongxiaorong}
    \vspace{-10pt}
\end{figure}

We also examine different mesh-driven deformation strategies in Fig.~\ref{fig:dongxiaorong}, comparing unbound Gaussians, rotation-based offset binding, and local affine deformation transfer. Unbound Gaussians lack mesh correspondence and cannot follow mesh manipulations. Offset binding transfers Gaussians through rotated vertex-relative offsets but does not account for local stretching and shearing, resulting in Gaussian--mesh misalignment and spike artifacts under large non-rigid deformations in \textit{Chair} and \textit{Drums}. In contrast, P-MGS updates Gaussian centers through barycentric coordinates and covariances through local affine transformations, preserving Gaussian--mesh alignment and producing cleaner, more coherent renderings.

\section{Conclusion}
In this work, we revisited the relationship between geometry and appearance in mesh--Gaussian splatting. We demonstrated that meshes provide geometric controllability, while mesh subdivision density does not reliably reflect appearance complexity. 
By decoupling appearance optimization from geometric control, P-MGS preserves adaptive Gaussian modeling and enables flexible editing. Experiments on synthetic and real-world datasets demonstrate improved novel-view synthesis quality with fewer Gaussian primitives and coherent mesh-driven manipulation, producing sharper renderings with fewer artifacts under deformation.
P-MGS offers a new perspective on designing editable Gaussian representations.

%% file: sec/X_suppl.tex
\clearpage
\setcounter{page}{1}
\maketitlesupplementary


\section{Additional Qualitative Comparisons}
\label{sec:additional_qualitative}

\begin{figure}[t]
    \centering
    \includegraphics[width=\columnwidth]{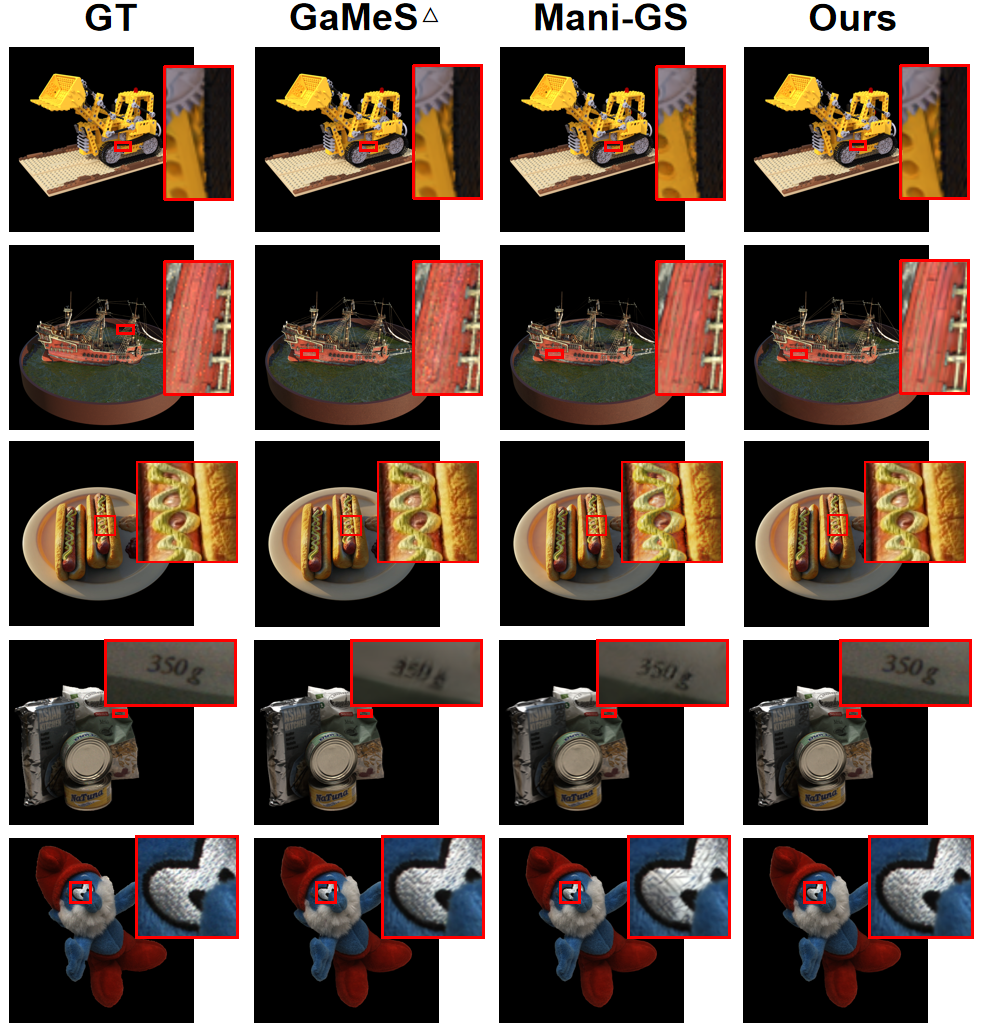}
    \caption{
Additional qualitative comparisons of static novel-view
rendering on \textit{Lego}, \textit{Ship},
and \textit{Hotdog} from NeRF-Synthetic, as well as
Scan97 and Scan83 from DTU.
    The zoomed-in regions highlight differences in
    texture and structural details.
    Compared with GaMeS$_{\triangle}$ and Mani-GS, P-MGS generally
    produces cleaner renderings with fewer artifacts
    and better preserves appearance details.
    }
    \label{fig:appendix_static}
    \vspace{-10pt}
\end{figure}

\begin{figure}[t]
    \centering
    \includegraphics[width=\columnwidth]{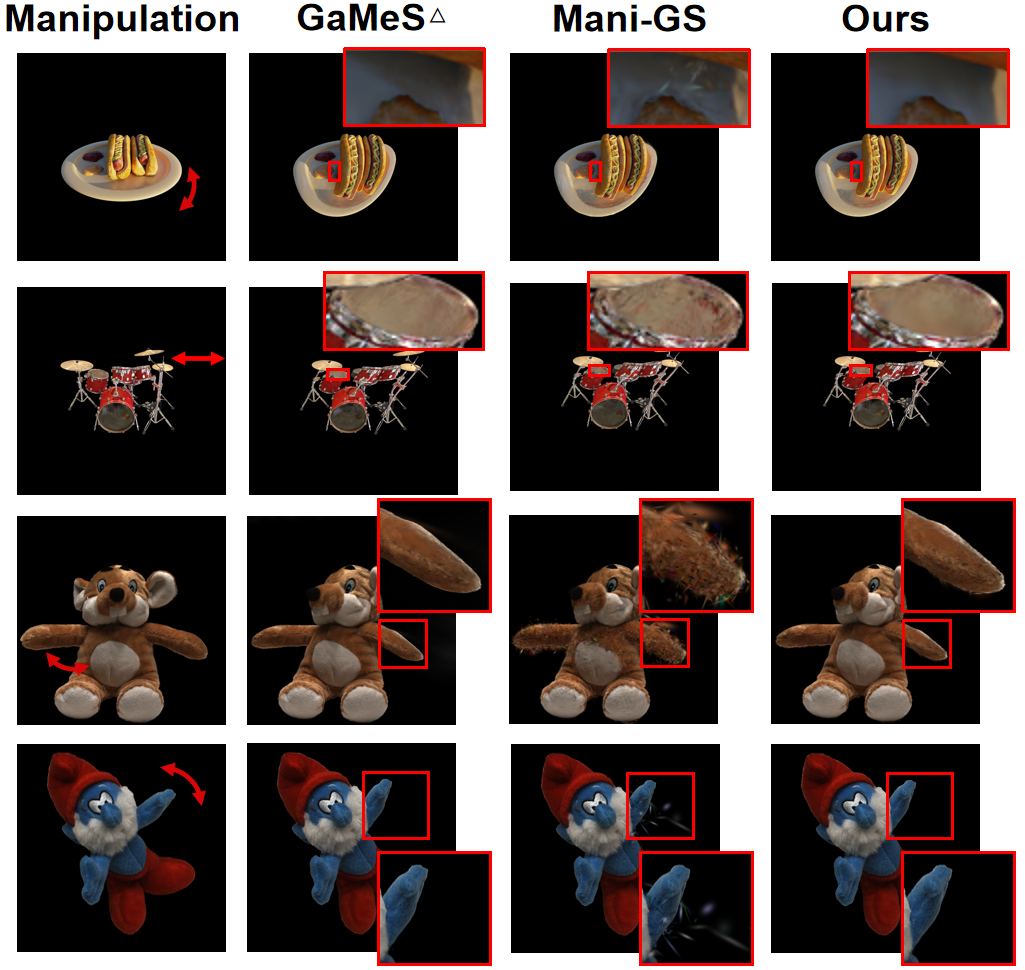}
    \caption{
    Additional qualitative comparisons of mesh-driven
    editing on \textit{Hotdog} and \textit{Drums} from
    NeRF-Synthetic, and Scan105 and Scan83 from DTU.
    Red arrows indicate the applied manipulations,
    and red boxes highlight the edited regions.
    GaMeS$_{\triangle}$ and Mani-GS exhibit floating Gaussians,
    rendering artifacts, holes, and Gaussian--mesh
    misalignment, particularly around manipulated regions,
    whereas P-MGS produces cleaner and more coherent
    rendering results.
    }
    \vspace{-10pt}
    \label{fig:appendix_manipulation}
\end{figure}

\textbf{Static Rendering.}
Figure~\ref{fig:appendix_static} further compares static rendering quality across synthetic and real-world scenes. GaMeS$_{\triangle}$ and Mani-GS exhibit local artifacts and loss of fine appearance details, whereas P-MGS produces cleaner renderings with better preservation of textures and structural boundaries.

\textbf{Manipulation Rendering.}
Figure~\ref{fig:appendix_manipulation} presents additional
mesh-driven editing comparisons on \textit{Hotdog} and
\textit{Drums} from NeRF-Synthetic, and Scan105 and Scan83
from DTU. GaMeS$_{\triangle}$ and Mani-GS exhibit floating Gaussians,
rendering artifacts, holes, and Gaussian--mesh misalignment,
particularly around manipulated regions. In contrast,
P-MGS maintains better Gaussian--mesh alignment,
more coherent object shapes, and cleaner renderings,
demonstrating the effectiveness of post-hoc binding
and local affine deformation transfer.

\section{Implementation Details}
\label{sec:allocation_implementation}

\subsection{Training Configuration}

All models are trained for 30,000 iterations on a single NVIDIA RTX 4090 GPU. Training time ranges from a few minutes to less than one hour, depending on scene complexity, dataset characteristics, Gaussian allocation, and mesh optimization settings.

\subsection{Observation Statistics and Allocation Bounds}

We set $\epsilon=(1/255)^2$ and estimate visible coverage using four subpixel rays per pixel. The foreground weight $a_v(p)$ is obtained from the original alpha channel of the training images.

Let $e_f\in\{0,1\}$ indicate whether face $f$ has nonzero area and passes training-frustum filtering. The face-level allocation bounds are defined as
\begin{equation}
\ell_f=e_f,\qquad
u_f=
\begin{cases}
0, & e_f=0,\\
1, & e_f=1,\ Q_f=0,\\
\max\!\left(1,\left\lceil\sum_v A_{fv}\right\rceil\right),
& e_f=1,\ Q_f>0.
\end{cases}
\label{eq:allocation_bounds}
\end{equation}
These bounds exclude ineligible faces, retain one Gaussian for eligible but unobserved faces, and limit allocation based on accumulated visible coverage.

The allocation is solved using marginal-threshold bisection for each fixed-price integer subproblem, followed by exact-budget completion. Nonnegative dual prices are optimized using L-BFGS-B with up to 80 iterations. We retain the workload-feasible assignment with the lowest surrogate error.

\subsection{Measured Workload Validation and Cost Correction}

We validate the allocated Gaussian configuration using rasterizer-reported tile counts. If the measured workload exceeds the reference limit $W_v$, we update the per-face cost estimates as
\begin{equation}
T_{fv}^{(r+1)}
=
\max\left(
T_{fv}^{(r)},
\max_{i\in\mathcal{I}_f^{(r)}} t_{iv}^{(r)}
\right),
\label{eq:workload_cost_correction}
\end{equation}
where $t_{iv}^{(r)}$ denotes the measured tile count of Gaussian $i$ in view $v$, and $\mathcal{I}_f^{(r)}$ contains the Gaussians initialized on face $f$ at correction round $r$. Estimates for faces without initialized Gaussians remain unchanged.

The allocation is updated for at most four correction rounds. The accepted initialization satisfies
\begin{equation}
\sum_i t_{iv}^{\mathrm{candidate}}\leq W_v,
\qquad \forall v.
\label{eq:actual_initialization_workload}
\end{equation}

\subsection{Post-hoc Gaussian--Mesh Binding}
\label{sec:binding_implementation}

Before post-hoc binding, we filter optimized Gaussians and reference mesh faces outside the predefined object bounds and remove zero-area mesh faces. This preprocessing excludes spatially detached primitives outside the object region and ensures valid geometric candidates for Gaussian--mesh correspondence.

We construct a KD-tree over the remaining mesh face centers and retrieve the 32 nearest candidate faces for each optimized Gaussian. The associated face is selected by minimizing the Euclidean distance from the Gaussian center to the closest point on each candidate triangle. The resulting face assignment remains fixed during deformation.

\subsection{Binding Coordinate Computation}
\label{sec:binding_affine_details}

For each Gaussian assigned to face $f_i$, we define
$\mathbf{A}_{f_i}^0=
[\mathbf{v}_{f_i,2}^0-\mathbf{v}_{f_i,1}^0,\,
\mathbf{v}_{f_i,3}^0-\mathbf{v}_{f_i,1}^0]$.
The signed normal offset and barycentric weights are computed as
\begin{equation}
\begin{aligned}
d_i &=
(\boldsymbol{\mu}_i^0-\mathbf{v}_{f_i,1}^0)^\top
\mathbf{n}_{f_i}^0,\\
\begin{bmatrix}
w_{i2}\\w_{i3}
\end{bmatrix}
&=
\left((\mathbf{A}_{f_i}^0)^\top
\mathbf{A}_{f_i}^0\right)^{-1}
(\mathbf{A}_{f_i}^0)^\top
\left(
\boldsymbol{\mu}_i^0
-d_i\mathbf{n}_{f_i}^0
-\mathbf{v}_{f_i,1}^0
\right),\\
w_{i1}&=1-w_{i2}-w_{i3}.
\end{aligned}
\end{equation}
The barycentric weights are not restricted to $[0,1]$,
preserving the optimized positions of Gaussians
projected outside their assigned triangles.

The resulting center deformation is equivalent to
\begin{equation}
\boldsymbol{\mu}_i^t
=
\mathbf{v}_{f_i,1}^t
+
\mathbf{J}_{f_i}^t
(\boldsymbol{\mu}_i^0-\mathbf{v}_{f_i,1}^0),
\end{equation}
confirming that Gaussian centers and covariances
follow the same local affine transformation.

\subsection{Identity Preservation of Post-hoc Binding}
\label{sec:binding_identity}

Post-hoc binding establishes Gaussian--mesh correspondence
while preserving the optimized Gaussian representation
in its reference configuration.
To isolate the effect of binding, we compare the same
filtered Gaussian set before and after coordinate
decomposition and reconstruction on undeformed reference meshes,
using the preprocessing described in
Sec.~\ref{sec:binding_implementation}.

Let $N$ denote the number of filtered Gaussians, and let
$\hat{\mu}_i^0 = \sum_{k=1}^{3} w_{ik} v_{f_i,k}^0 + d_i n_{f_i}^0$
denote the center reconstructed from the bound coordinates in Eq.~(7).
We evaluate the PSNR difference and the mean and
maximum Gaussian center reconstruction errors:
\begin{equation}
\begin{aligned}
\Delta_{\mathrm{bind}}
&=
\mathrm{PSNR}_{\mathrm{after}}
-
\mathrm{PSNR}_{\mathrm{before}},\\
e_{\mu}
&=
\frac{1}{N}
\sum_{i=1}^{N}
\left\|
\hat{\mu}_i^0
-
\mu_i^0
\right\|_2,\\
e_{\mu}^{\max}
&=
\max_i
\left\|
\hat{\mu}_i^0
-
\mu_i^0
\right\|_2.
\end{aligned}
\end{equation}

We perform this evaluation on all eight NeRF-Synthetic
scenes using 200 test views per scene at $800\times800$
resolution. Both rendering passes use identical cameras
and black backgrounds.

\begin{table}[t]
\centering
\small
\caption{
Identity preservation of post-hoc binding
on undeformed reference meshes.
Comparisons use the same filtered Gaussian set
before and after binding.
Rendering quality and Gaussian centers are
preserved to numerical precision across
all eight NeRF-Synthetic scenes.
}
\label{tab:binding-identity}

\renewcommand{\arraystretch}{1.1}
\setlength{\tabcolsep}{4pt}

\resizebox{\columnwidth}{!}{
\begin{tabular}{lccc}
\toprule
\textbf{Scene}
& $\boldsymbol{\Delta_{\rm bind}}$ \textbf{(dB)}
& $\boldsymbol{e_\mu}$
& $\boldsymbol{e_\mu^{\max}}$ \\
\midrule

Chair
& $-1.038\times10^{-8}$
& $6.281\times10^{-17}$
& $2.152\times10^{-15}$ \\

Drums
& $-3.318\times10^{-8}$
& $6.909\times10^{-17}$
& $6.431\times10^{-13}$ \\

Ficus
& $1.496\times10^{-8}$
& $5.183\times10^{-17}$
& $2.334\times10^{-14}$ \\

Hotdog
& $2.904\times10^{-7}$
& $5.028\times10^{-17}$
& $2.121\times10^{-15}$ \\

Lego
& $1.749\times10^{-7}$
& $5.402\times10^{-17}$
& $1.322\times10^{-14}$ \\

Materials
& $1.932\times10^{-7}$
& $5.285\times10^{-17}$
& $3.868\times10^{-15}$ \\

Mic
& $2.258\times10^{-7}$
& $8.631\times10^{-17}$
& $3.360\times10^{-14}$ \\

Ship
& $-2.369\times10^{-7}$
& $7.545\times10^{-17}$
& $6.824\times10^{-13}$ \\

\midrule
\textbf{Overall}
& $7.734\times10^{-8}$
& $6.283\times10^{-17}$
& $6.824\times10^{-13}$ \\

\bottomrule
\end{tabular}
}
\vspace{-8pt}
\end{table}

As shown in Table~\ref{tab:binding-identity},
post-hoc binding preserves the filtered Gaussian
representation to numerical precision across
all eight NeRF-Synthetic scenes.
The average PSNR difference before and after binding
is $7.734\times10^{-8}$ dB, while the mean Gaussian
center reconstruction error is $6.283\times10^{-17}$,
with a maximum error of $6.824\times10^{-13}$.
These results confirm that post-hoc binding establishes
explicit Gaussian--mesh correspondence without introducing
additional reconstruction errors.

\subsection{View-Dependent Appearance under Mesh Deformation}
\label{sec:sh_appearance}

To transport view-dependent appearance during mesh-driven editing,
we retain the reference-pose SH coefficients and inverse-rotate
the viewing direction:
\begin{equation}
\mathbf{C}_{i,\mathrm{rot}}^t
=
\left[
\sum_{\ell=0}^{L}\sum_{m=-\ell}^{\ell}
\mathbf{b}_{i,\ell m}^{0}
Y_{\ell m}\!\left(
(\mathbf{R}_{f_i}^t)^\top
\boldsymbol{\omega}_{o,i}^{t}
\right)
\right]_+,
\label{eq:sh_deformation}
\end{equation}
where $\mathbf{R}_{f_i}^t$ is the proper polar rotation of
$\mathbf{J}_{f_i}^t$, and $\boldsymbol{\omega}_{o,i}^{t}$ points
from the deformed Gaussian towards the camera. The fixed coefficients
use this outgoing-direction convention, with the constant color
offset absorbed into the DC term.

\begin{figure}[t]
\centering
\includegraphics[width=\linewidth]{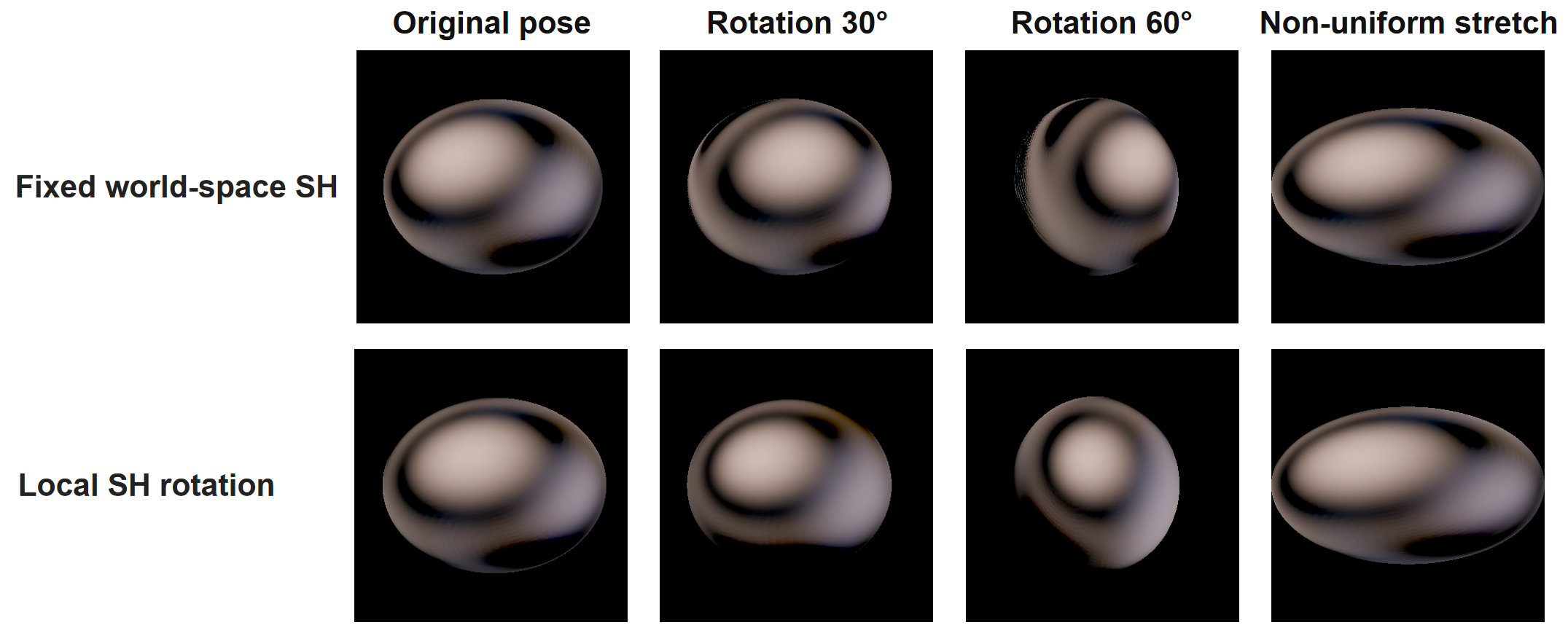}
\caption{
Controlled surface-shading illustration of SH appearance transport.
Columns show the original pose, rotations of $30^\circ$ and $60^\circ$,
and non-uniform stretch. Top: fixed world-space SH evaluated at
$\boldsymbol{\omega}_{o,i}^{t}$.
Bottom: local SH rotation using Eq.~\eqref{eq:sh_deformation}.
Both rows share the geometry, camera, and display transform and use
the same degree-three SH coefficients fitted in the reference pose.
}
\label{fig:sh_fixed_world}
\end{figure}

Figure~\ref{fig:sh_fixed_world} compares direct world-space SH
evaluation with local SH rotation. Both rows coincide in the original
pose, where $\mathbf{R}_{f_i}^0=\mathbf{I}$. After deformation,
local rotation transports the reference directional appearance with
the mesh without retraining. This operation transports baked radiance;
it does not generally recompute reflections under fixed world illumination.

\section{Future Work}

P-MGS currently focuses on reconstructing editable 3D assets from multi-view observations. In future work, we aim to extend this framework to single-image 3D asset generation, enabling high-quality appearance synthesis while preserving mesh-driven geometric controllability.

%% file: main.bib
@String(CVPR= {IEEE Conf. Comput. Vis. Pattern Recog.})

@String(ICCV= {Int. Conf. Comput. Vis.})

@String(ECCV= {Eur. Conf. Comput. Vis.})

@String(TIP  = {IEEE Trans. Image Process.})

@String(AAAI = {AAAI})

@String(CVPR  = {CVPR})

@String(ICCV  = {ICCV})

@String(ECCV  = {ECCV})

@String(TIP   = {IEEE TIP})

@article{3dgs,
author = {Kerbl, Bernhard and Kopanas, Georgios and Leimkuehler, Thomas and Drettakis, George},
title = {3D Gaussian Splatting for Real-Time Radiance Field Rendering},
year = {2023},
volume = {42},
number = {4},
journal = {ACM Trans. Graph.},
articleno = {139},
numpages = {14},
}

@article{mildenhall,
  title={Nerf: Representing scenes as neural radiance fields for view synthesis},
  author={Mildenhall, Ben and Srinivasan, Pratul P. and Tancik, Matthew and Barron, Jonathan T. and Ramamoorthi, Ravi and Ng, Ren},
  journal={Communications of the ACM},
  volume={65},
  number={1},
  pages={99--106},
  year={2021},
  publisher={ACM, New York, NY, USA}
}

@InProceedings{colmap,
author = {Schonberger, Johannes L. and Frahm, Jan-Michael},
title = {Structure-From-Motion Revisited},
booktitle = CVPR,
year = {2016}
}

@inproceedings{plenoxels,
  title={Plenoxels: Radiance fields without neural networks},
  author={Fridovich-Keil, Sara and Yu, Alex and Tancik, Matthew and Chen, Qinhong and Recht, Benjamin and Kanazawa, Angjoo},
  booktitle = CVPR,
  pages={5491--5500},
  year={2022},
  organization={IEEE}
}

@inproceedings{nerfacc,
  title={Nerfacc: Efficient sampling accelerates nerfs},
  author={Li, Ruilong and Gao, Hang and Tancik, Matthew and Kanazawa, Angjoo},
  booktitle = ICCV,
  pages={18491--18500},
  year={2023},
  organization={IEEE}
}

@inproceedings{sugar,
  title={Sugar: Surface-aligned gaussian splatting for efficient 3d mesh reconstruction and high-quality mesh rendering},
  author={Gu{\'e}don, Antoine and Lepetit, Vincent},
  booktitle = CVPR,
  pages={5354--5363},
  year={2024},
  organization={IEEE}
}

@article{games,
  title={Games: Mesh-based adapting and modification of gaussian splatting},
  author = {Joanna Waczyńska and Piotr Borycki and Sławomir Tadeja and Jacek Tabor and Przemysław Spurek},
  journal = {Pattern Recognition},
  volume = {181},
  pages={114561},
  year = {2027},
  issn = {0031-3203},
}

@article{gaomesh,
  title={Mesh-based gaussian splatting for real-time large-scale deformation},
  author={Gao, Lin and Yang, Jie and Zhang, Bo-Tao and Sun, Jia-Mu and Yuan, Yu-Jie and Fu, Hongbo and Lai, Yu-Kun},
  journal={arXiv preprint arXiv:2402.04796},
  year={2024}
}

@inproceedings{manigs,
  title={Mani-gs: Gaussian splatting manipulation with triangular mesh},
  author={Gao, Xiangjun and Li, Xiaoyu and Zhuang, Yiyu and Zhang, Qi and Hu, Wenbo and Zhang, Chaopeng and Yao, Yao and Shan, Ying and Quan, Long},
   booktitle = CVPR,
  pages={21392--21402},
  year={2025},
  organization={IEEE}
}

@InProceedings{meshsplatting,
    author    = {Held, Jan and Son, Sanghyun and Vandeghen, Renaud and Rebain, Daniel and Gadelha, Matheus and Zhou, Yi and Cioppa, Anthony and Lin, Ming C. and Van Droogenbroeck, Marc and Tagliasacchi, Andrea},
    title     = {MeshSplatting: Differentiable Rendering with Opaque Meshes},
    booktitle = {Proceedings of the IEEE/CVF Conference on Computer Vision and Pattern Recognition (CVPR)},
    month     = {June},
    year      = {2026},
    pages     = {7320-7329}
}

@inproceedings{yu,
  title={Mip-splatting: Alias-free 3d gaussian splatting},
  author={Yu, Zehao and Chen, Anpei and Huang, Binbin and Sattler, Torsten and Geiger, Andreas},
  booktitle = CVPR,
  pages={19447--19456},
  year={2024},
  organization={IEEE}
}

@inproceedings{wu,
  title={4d gaussian splatting for real-time dynamic scene rendering},
  author={Wu, Guanjun and Yi, Taoran and Fang, Jiemin and Xie, Lingxi and Zhang, Xiaopeng and Wei, Wei and Liu, Wenyu and Tian, Qi and Wang, Xinggang},
  booktitle = CVPR,
  pages={20310--20320},
  year={2024},
  organization={IEEE}
}

@inproceedings{jiang,
  title={Gaussianshader: 3d gaussian splatting with shading functions for reflective surfaces},
  author={Jiang, Yingwenqi and Tu, Jiadong and Liu, Yuan and Gao, Xifeng and Long, Xiaoxiao and Wang, Wenping and Ma, Yuexin},
  booktitle = CVPR,
  pages={5322--5332},
  year={2024},
  organization={IEEE}
}

@inproceedings{dtu,
  title={Large scale multi-view stereopsis evaluation},
  author={Jensen, Rasmus and Dahl, Anders and Vogiatzis, George and Tola, Engin and Aan{\ae}s, Henrik},
  booktitle = CVPR,
  pages={406--413},
  year={2014}
}

@article{ssim,
  title={Image quality assessment: from error visibility to structural similarity},
  author={Wang, Zhou and Bovik, Alan C and Sheikh, Hamid R and Simoncelli, Eero P},
  journal = TIP,
  volume={13},
  number={4},
  pages={600--612},
  year={2004},
  publisher={IEEE}
}

@inproceedings{lpips,
  title={The unreasonable effectiveness of deep features as a perceptual metric},
  author={Zhang, Richard and Isola, Phillip and Efros, Alexei A and Shechtman, Eli and Wang, Oliver},
  booktitle = CVPR,
  pages={586--595},
  year={2018},
  organization={IEEE}
}

@inproceedings{nerfediting,
  title={Nerf-editing: geometry editing of neural radiance fields},
  author={Yuan, Yu-Jie and Sun, Yang-Tian and Lai, Yu-Kun and Ma, Yuewen and Jia, Rongfei and Gao, Lin},
  booktitle = CVPR,
  pages={18332--18343},
  year={2022},
  organization={IEEE}
}

@inproceedings{neumesh,
  title={Neumesh: Learning disentangled neural mesh-based implicit field for geometry and texture editing},
  author={Yang, Bangbang and Bao, Chong and Zeng, Junyi and Bao, Hujun and Zhang, Yinda and Cui, Zhaopeng and Zhang, Guofeng},
  booktitle = ECCV,
  pages={597--614},
  year={2022},
  organization={Springer}
}

@inproceedings{neuralimpostor,
  title={Neural impostor: Editing neural radiance fields with explicit shape manipulation},
  author={Liu, Ruiyang and Xiang, Jinxu and Zhao, Bowen and Zhang, Ran and Yu, Jingyi and Zheng, Changxi},
  booktitle={Computer Graphics Forum},
  volume={42},
  number={7},
  pages={e14981},
  year={2023},
  organization={Wiley Online Library}
}

@article{nerfshop,
  title={Nerfshop: Interactive editing of neural radiance fields},
  author={Jambon, Cl{\'e}ment and Kerbl, Bernhard and Kopanas, Georgios and Diolatzis, Stavros and Leimk{\"u}hler, Thomas and Drettakis, George},
  journal={Proceedings of the ACM on Computer Graphics and Interactive Techniques},
  volume={6},
  number={1},
  year={2023}
}

@inproceedings{mobilenerf,
  title={Mobilenerf: Exploiting the polygon rasterization pipeline for efficient neural field rendering on mobile architectures},
  author={Chen, Zhiqin and Funkhouser, Thomas and Hedman, Peter and Tagliasacchi, Andrea},
  booktitle = CVPR,
  pages={16569--16578},
  year={2023},
  organization={IEEE}
}

@inproceedings{gaussianfrosting,
  title={Gaussian frosting: Editable complex radiance fields with real-time rendering},
  author={Gu{\'e}don, Antoine and Lepetit, Vincent},
  booktitle = ECCV,
  pages={413--430},
  year={2024},
  organization={Springer}
}

@inproceedings{scgs,
  title={Sc-gs: Sparse-controlled gaussian splatting for editable dynamic scenes},
  author={Huang, Yi-Hua and Sun, Yang-Tian and Yang, Ziyi and Lyu, Xiaoyang and Cao, Yan-Pei and Qi, Xiaojuan},
  booktitle = CVPR,
  pages={4220--4230},
  year={2024},
  organization={IEEE}
}

@inproceedings{physgaussian,
  title={Physgaussian: Physics-integrated 3d gaussians for generative dynamics},
  author={Xie, Tianyi and Zong, Zeshun and Qiu, Yuxing and Li, Xuan and Feng, Yutao and Yang, Yin and Jiang, Chenfanfu},
  booktitle = CVPR,
  pages={4389--4398},
  year={2024},
  organization={IEEE}
}

@inproceedings{vrgs,
author = {Jiang, Ying and Yu, Chang and Xie, Tianyi and Li, Xuan and Feng, Yutao and Wang, Huamin and Li, Minchen and Lau, Henry and Gao, Feng and Yang, Yin and Jiang, Chenfanfu},
title = {VR-GS: A Physical Dynamics-Aware Interactive Gaussian Splatting System in Virtual Reality},
year = {2024},
isbn = {9798400705250},
booktitle = {ACM SIGGRAPH 2024 Conference Papers},
articleno = {78},
numpages = {1},
}

@article{neus,
  title={Neus: Learning neural implicit surfaces by volume rendering for multi-view reconstruction},
  author={Wang, Peng and Liu, Lingjie and Liu, Yuan and Theobalt, Christian and Komura, Taku and Wang, Wenping},
  journal={arXiv preprint arXiv:2106.10689},
  year={2021}
}

@inproceedings{unimgs,
  title={UniMGS: Unifying Mesh and 3D Gaussian Splatting with Single-Pass Rasterization and Proxy-Based Deformation},
  author={Xiao, Zeyu and Sun, Mingyang and Cong, Yimin and Wang, Lintao and Kou, Dongliang and Wu, Zhenyi and Yang, Dingkang and Zhai, Peng and Wang, Zeyu and Zhang, Lihua},
  booktitle={Proceedings of the AAAI Conference on Artificial Intelligence},
  volume={40},
  number={13},
  pages={11023--11031},
  year={2026}
}
